\documentclass[a4paper,11pt]{article}
\pdfoutput=1 

\usepackage{jcappub} 

\usepackage[T1]{fontenc} 

\title{\boldmath Properties of white dwarfs in Einstein-$\Lambda$ gravity}

\author[a,1]{H.L. Liu,\note{Corresponding author.}}
\author[a]{G.L. L\"{u}}

\affiliation[a]{School of Physical Science and Technology, Xinjiang University,\\Urumuqi, 830046, China}

\emailAdd{heleiliu@xju.edu.cn}

\abstract{In this paper, we explore the properties of white dwarfs with the modified TOV equation in Einstein-$\Lambda$ gravity, the equilibrium configurations predict a maximum mass limit for white dwarfs same as the Chandrasekhar limit when we consider $\Lambda$ to be very small($\Lambda<10^{-16}$ m$^{-2}$),  by increasing $\Lambda$, the maximum mass and radius of white dwarf are reduced.
We study effects of the cosmological constant on the physical properties of white dwarf such as $M-\rho_c$ relation, $M-R$ relation, Schwarzschild radius, average density, compactness, gravitational redshift and dynamical stability. The gravitational redshift is a decreasing function of cosmological constant, because the gravitational redshift of white dwarf should be positive, we also find an upper limit for $\Lambda$, namely, $\Lambda<3\times10^{-14}$ m$^{-2}$. Our investigation of dynamical stability shown that the white dwarfs follow the dynamical stability in Einstein-$\Lambda$ gravity.}

\begin{document}
\maketitle
\flushbottom

\section{Introduction}
\label{sec:intro}

 Tolman, Oppenheimer, and Volkoff(TOV) made the first attempt for obtaining Einsteinian hydrostatic equilibrium equation for stars~\cite{Tolman1,Tolman2,Oppenheimer},
the physical characteristics of stars using TOV equation have been investigated by many authors~\cite{silbar,narain,bordbar,li,oliveira,liu,wen}. Recently, the study of modified gravity has attracted many researchers's attention, a large number of modified theories of gravity
 such as $F(R)$~\cite{astashenok,momeni}, $F(G)$~\cite{abbas}, dilaton gravities~\cite{hendi}, gravity's rainbow~\cite{a}, Einstein-$\Lambda$ gravity~\cite{Bordbar2016}, massive gravity~\cite{kats} and post-Newtonian theory~\cite{glam} have been proposed.
 Compact objects provide an arena for understanding the properties of modified gravity that differ from Einstein's general relativity, because the gravitational field of white dwarf is much weaker than neutron star, the modified TOV equation in gravity's theory has been applied to neutron stars~\cite{astashenok,momeni,abbas,hendi,a,Bordbar2016,kats,glam,astashenok13,orellana}, but has rarely been explored for white dwarfs~\cite{das1,das,Jain,carvalho}.

 In this work, we aim at exploring the effect of modified TOV equation in Einstein-$\Lambda$ gravity in white dwarfs, and expect to see interesting features of these modifications inside stellar objects.

The plan of this paper is as follows. First, we briefly review the 4-dimensional hydrostatic equilibrium equation in Einstein-$\Lambda$ gravity, then we recall the equation of state (EOS) and boundary conditions. In section~\ref{property}, we present the structure properties of white dwarf in Einstein-$\Lambda$ gravity. We conclude with a short summary in the last section.

\section{4-dimensional hydrostatic equilibrium equation in Einstein-$\Lambda$ gravity}
To get the properties of white dwarfs in Einstein-$\Lambda$ gravity, we introduce the hydrostatic equilibrium equation in Einstein-$\Lambda$ gravity firstly~\cite{a,Bordbar2016}.
The action of Einstein gravity with the cosmological constant in 4 dimensions is given by:
\begin{equation}
\label{IG}
I_{G}=-\frac{1}{16\pi}\int (R-2\Lambda)\sqrt{-g}d^{4}x+I_{\rm Matt},
\end {equation}
where $g$ is the determinant of the metric tensor, $R$ is the Ricci scalar which is defined by $R=g^{\mu\nu}R_{\mu\nu}$, $\Lambda$ is the cosmological constant, $I_{\rm Matt}$ is the action of matter field. Varying the action \eqref{IG} with respect to $g_\nu^\mu$, one obtains the equation of motion for this gravity:

\begin{equation}\label{Ee}
G_\mu^\nu+\Lambda g_\mu^\nu=KT_\mu^\nu,
\end{equation}
where $K=\frac{8\pi G}{c^{4}}$, $G_\mu^\nu$ and $T_\mu^\nu$ are the symmetric Ricci tensor and energy-momentum tensors of the matter field, respectively.

Assume a spherical symmetric metric in Einstein-$\Lambda$ gravity as
\begin{equation}\label{metric}
ds^{2}=f(r)dt^{2}-\frac{dr^2}{g(r)}-r^2(d\theta ^{2}+sin^2\theta d\varphi ^2),
\end{equation}

The energy-momentum tensor in the local frame is
\begin{equation}\label{EM}
T^{\mu\nu}=-Pg^{\mu\nu}+(c^2\rho+P)U^\mu U^\nu,
\end{equation}
where $P$ and $\rho$ are pressure and density of the fluid, respectively, $U^\mu$ is the four-velocity.

Using eqs.~\eqref{Ee} and~\eqref{metric}, the components of energy-momentum for 4-dimensions as follows
\begin{equation}\label{components}
T_0^{0}=\rho c^2 \qquad \& \qquad T_{1}^{1}=T_{2}^{2}=T_{3}^{3}=-P.
\end{equation}

Considering the metric eq.~\eqref{metric} and eq.~\eqref{components} for perfect fluid, we can obtain the components of eq. ~\eqref{Ee} with the following forms
\begin{align}
Kc^2r^2\rho&=\Lambda r^2+(1-g)-rg' \label{com1},\\
Kr^2fP&=-\Lambda r^2f-(1-g)f+rgf' \label{com2},\\
4Krf^2P&=-4\Lambda rf^2+2(gf)'f+r(g'f'+2gf'')f-rgf'^2,\label{com3}
\end{align}
where $f$, $g$, $\rho$ and $P$ are functions of $r$, the prime and double prime denote the first and second derivatives with respect to $r$, respectively.

Using Eqs.~\eqref{com1}-\eqref{com3}, and after some calculations, we obtain
\begin{equation}\label{phee}
\frac{dP}{dr}+\frac{f'}{2f}(c^2\rho+P)=0.
\end{equation}

We get $f'$ from eq.~\eqref{com2} as follow
\begin{equation}\label{fp}
f'=\frac{[r^2(\Lambda+KP)+(1-g)]f}{rg}.
\end{equation}

Based on eq.~\eqref{com1}, $g$ can be given by
\begin{equation}\label{gp}
g=1+\frac{\Lambda}{3}r^2-\frac{c^2KM}{4\pi r},
\end{equation}
where $M$ is a function of $r$: $M=\int 4\pi r^2\rho(r)dr$. By inserting eqs. ~\eqref{fp}and\eqref{gp} in eq.~\eqref{phee}, we get the the hydrostatic equilibrium in Einstein-$\Lambda$ gravity as
\begin{equation}\label{HEE}
 \frac{dP}{dr}=(c^{2}\rho+P)\frac{[3c^{2}GM+r^{3}(\Lambda c^{4}+12\pi GP)]}{c^{2}r[6GM-c^{2}r(\Lambda r^{2}+3)]}.
\end{equation}
For $\Lambda=0$, the usual TOV equation will be recovered perfectly~\cite{Tolman1,Tolman2,Oppenheimer}.

As it was shown, the obtained hydrostatic equilibrium equation in Einstein-$\Lambda$ gravity depends to the cosmological constant $\Lambda$.

\section{Equation of state and boundary conditions}
\begin{figure}
\centering 
\includegraphics [width=0.8\textwidth]{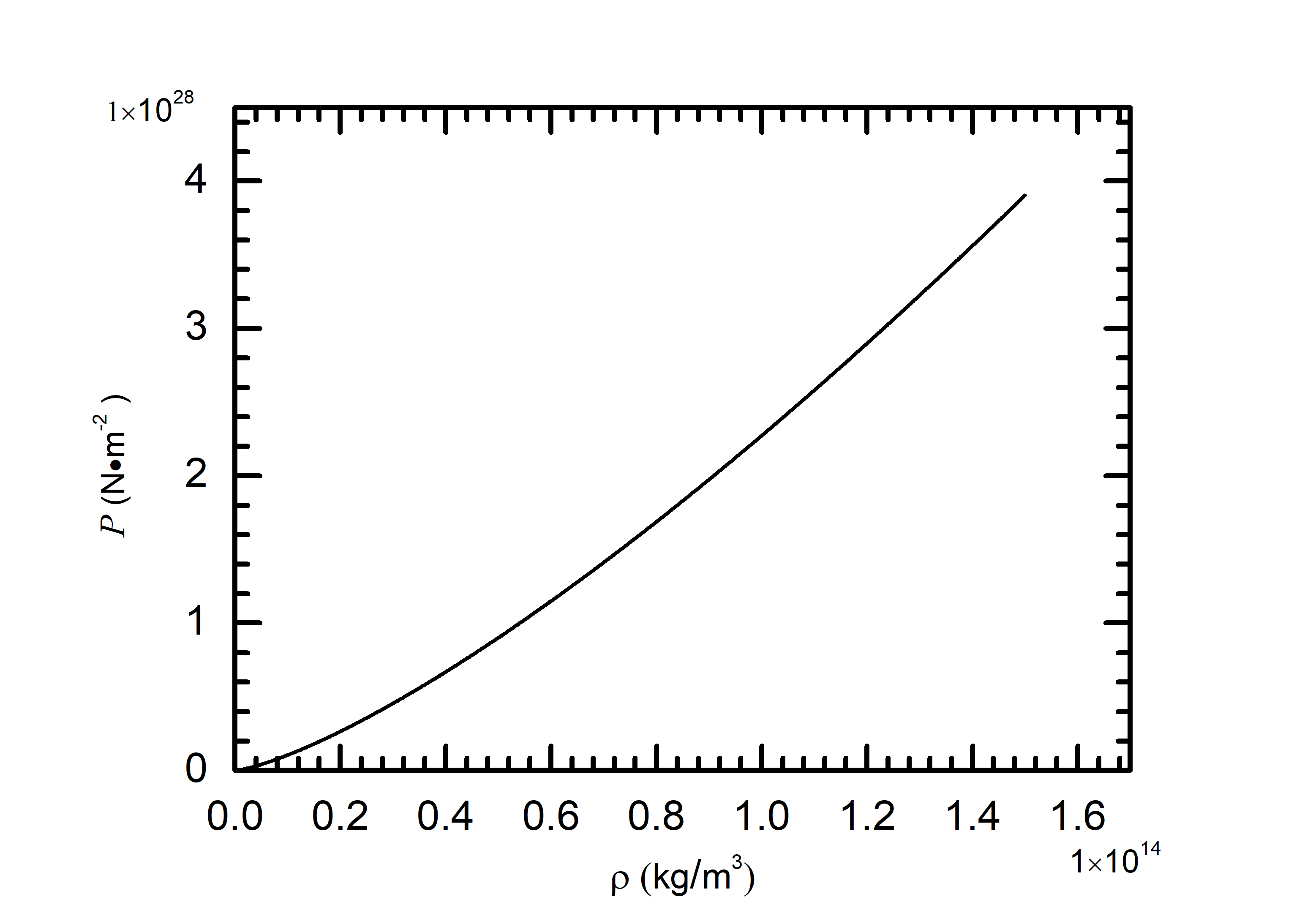}

\caption{\label{eos} Chandrasekhar's equation of state.}
\end{figure}

We use the equation of state(EOS) of white dwarfs obtained by Chandrasekhar~\cite{Ch1935}, which are constituted of electron degenerate matter,
\begin{equation}
P=\frac{8\pi c}{3(2\pi\hbar)^3}\int_{0}^{k_F}\frac{k^{2}}{(k^{2}+m_e^2c^2)^{1/2}}k^2dk,
\end{equation}
and
\begin{equation}
k_F=\hbar(\frac{3\pi^2\rho}{m_p\mu})^{1/3}.
\end{equation}
where $\hbar=h/{2\pi}$, $h$ is the Plank's constant, $k$ is the momentum of electrons, $\mu_e$ is the mean molecular weight per electron (we choose $\mu_e=2$ for our work), $m_p$ is the mass of a proton. The Chandrasekhar's EoS of the electron degenerate matter was shown in Fig.~\ref{eos}.

\begin{figure}[tbp]
\centering 
\includegraphics[width=.45\textwidth]{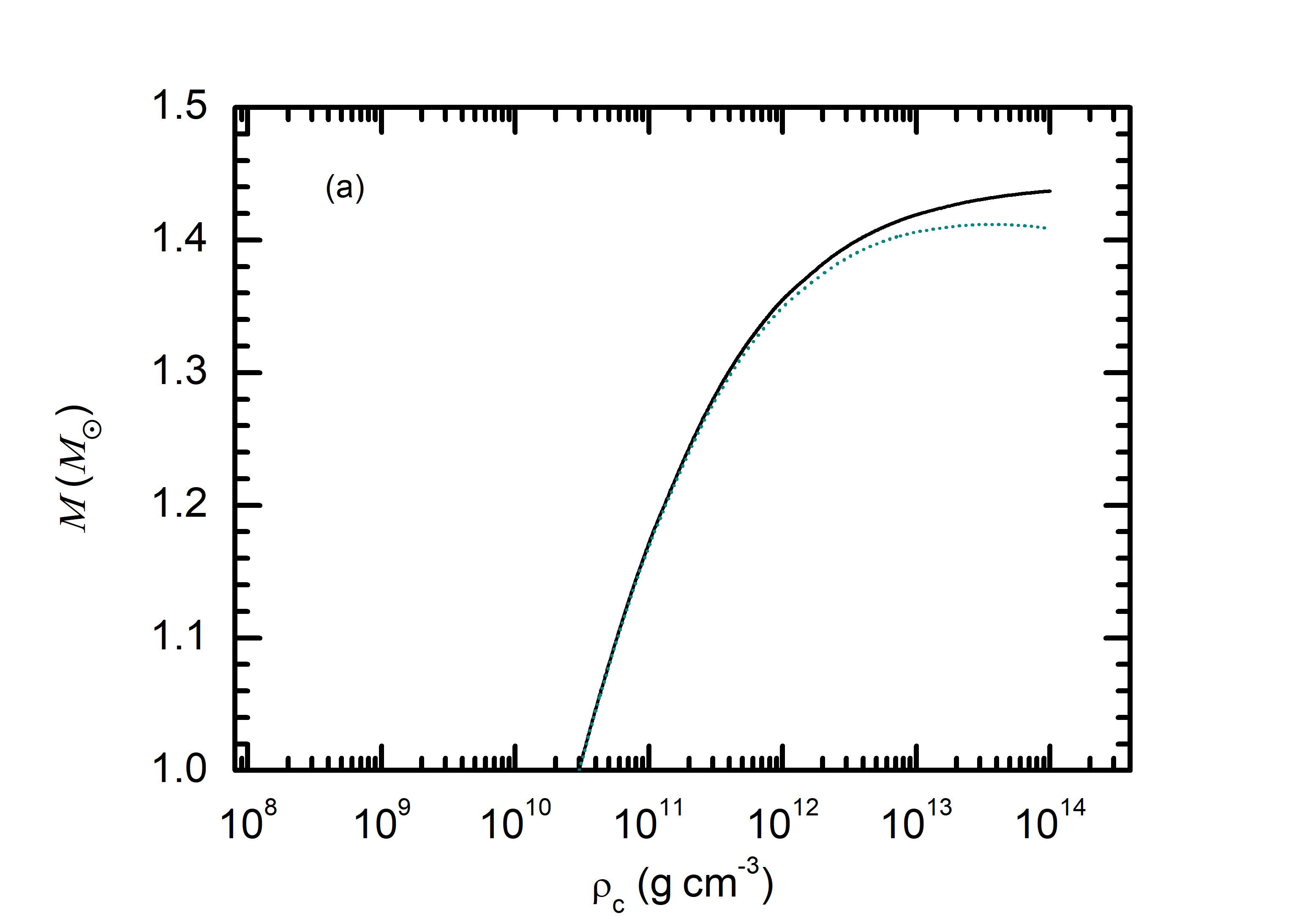}
\hfill
\includegraphics[width=.45\textwidth]{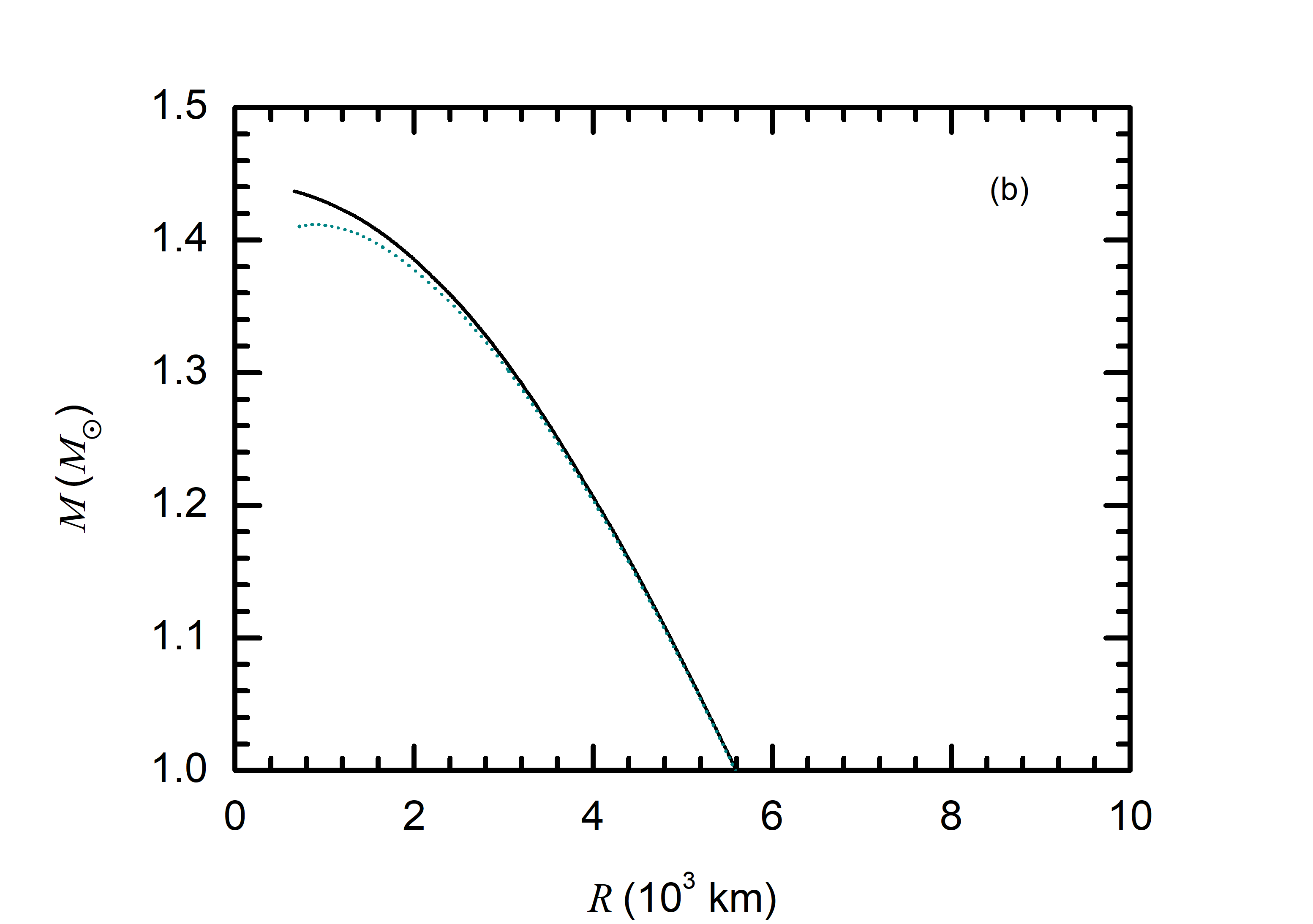}
\caption{\label{MRnt} Comparison of solutions in the Newtonian case (solid lines) and the general relativistic case (dotted lines). (a) Variation of $\rho_c$ with $M$, (b) $M-R$ relations.}
\end{figure}

The hydrostatic equilibrium equation in Einstein-$\Lambda$ gravity, accompanied by the above EoS, can be solved numerically with boundary conditions $M(r=0)=0$ and $\rho(r=0)=\rho_c$, where $\rho_c$ is the central density of the white dwarf.

In figure~\ref{MRnt}, we compare the Newtonian solutions with those in general relativity(non-modified Einstein gravity), i.e., $\Lambda=0$ case. Figure~\ref{MRnt} confirms that in the Newtonian case, with the increase of $\rho_c$, $M$ increases and $R$ decreases, until it saturates to a maximum mass $M_{max}\sim1.44M_{\odot}$, which is the famous Chandrasekhar limit. We further confirm that in the general relativistic case, with the increase of $\rho_c$, $M$ increases and $R$ decreases until it reaches $M_{max}=1.41M_{\odot}$ at $\rho_c=3.6\times10^{13}$ kg/m$^{3}$. A further increase in $\rho_c$ results in a slight decrease in $M$, indicating the onset of an unstable branch, which is absent in the Newtonian case. For low density white dwarfs having $\rho_c<10^{11}$ kg/m$^3$, the Newtonian and general relativistic $M-\rho_c$ curves are identical. However, for $ \rho_c>10^{11}~\rm kg/m^3$, general relativistic effects become important, leading to a slightly smaller $M$ compared to the Newtonian case, as a result, resulting in a smaller $M_{\rm max}$.

\section{Structure properties of white dwarf in Einstein-$\Lambda$ gravity}\label{property}
In this part, we will show the results obtained from our calculations. So far, the value of the cosmology constant is still an open question, it allowed us to consider it as a free parameter. The effects of cosmological constant on neutron stars have been investigated~\cite{Bordbar2016}, as the negative values of the cosmological constant are not logical for explaining the structure of neutron star. In this work, we only take the effects of variation of the positive cosmological constant into account to obtain the maximum mass of white dwarf by employing the Chandrasekhar's EOS. Although the $\Lambda$-effect in white dwarf has not yet been confirmed, it still can serve as an useful toy model, especially in the theoretical discussion.

\begin{figure}[tbp]
\centering 
\includegraphics[width=.65\textwidth]{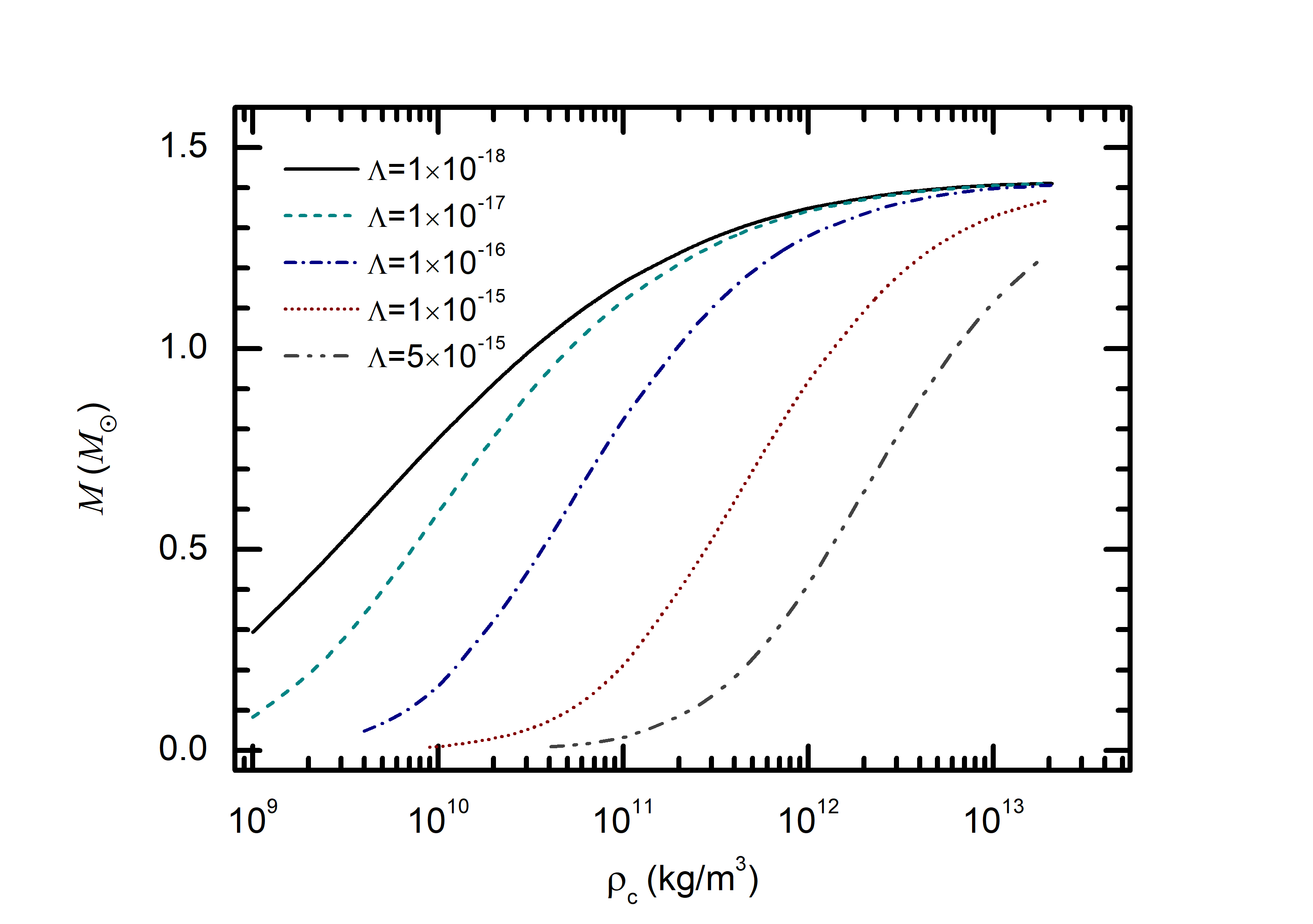}
\caption{\label{Mrho} Gravitational mass of white dwarf versus central density, different line represents different $\Lambda$, $\Lambda=1\times10^{-18} \rm (solid), 1\times10^{-17} \rm (dash), 1\times10^{-16}$(dash dot), $\Lambda=1\times10^{-15}$(dot) and $\Lambda=5\times10^{-15}$(dash dot dot).}
\end{figure}

\begin{figure}[tbp]
\centering 
\includegraphics[width=.65\textwidth]{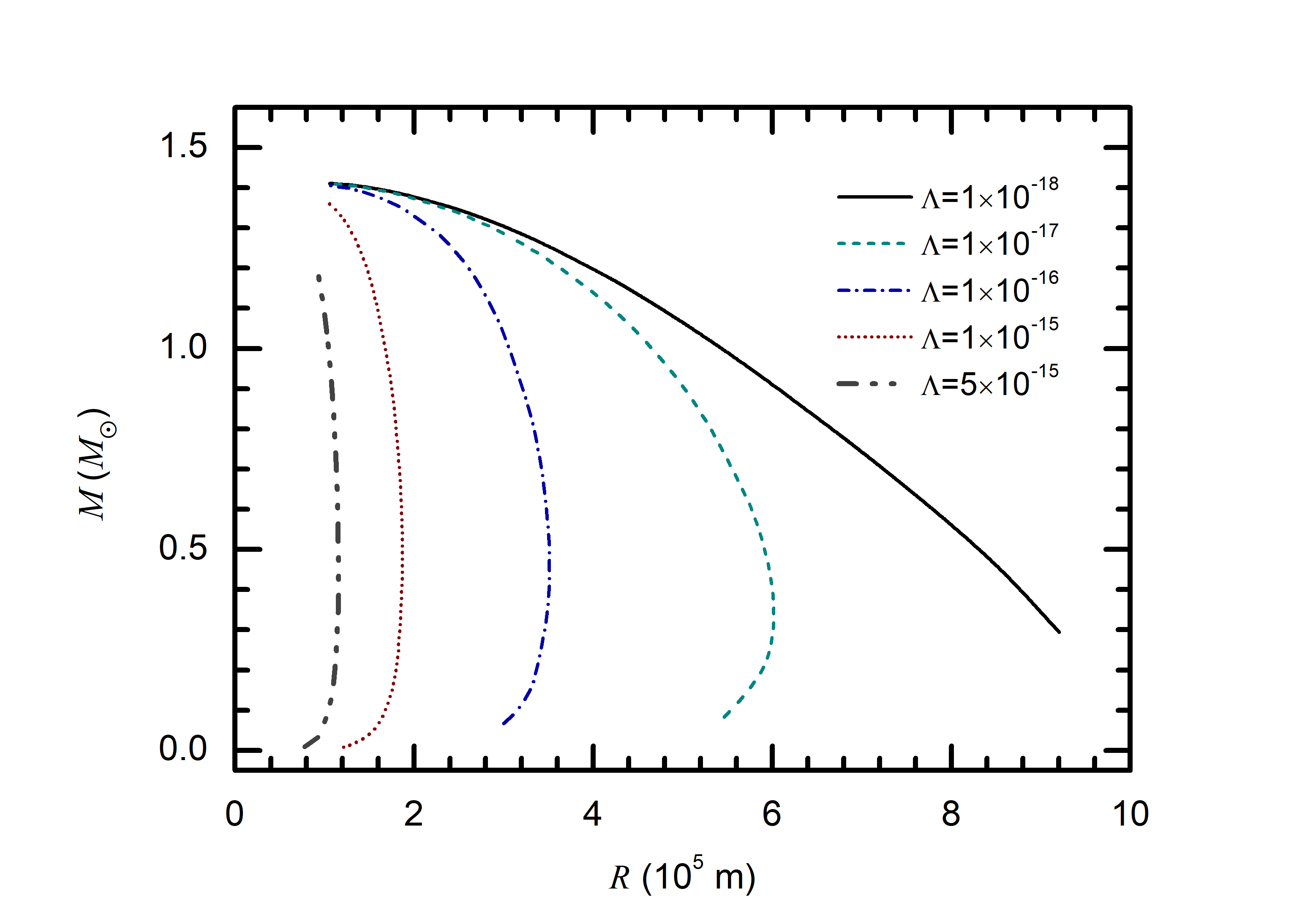}
\caption{\label{lmr} Gravitational mass of white dwarf versus radius for  $\Lambda=1\times10^{-18}$(solid), $ \Lambda=1\times10^{-17}$(dash), $\Lambda=1\times10^{-16}$(dash dot), $\Lambda=1\times10^{-15}$(dot), $\Lambda=5\times10^{-15}$(dash dot dot), respectively.}
\end{figure}

Our results are presented in table~\ref{lstructure} which shows the maximum mass of white dwarf with different cosmological constant $\Lambda$. It's notable that, the maximum mass of white dwarf reduces to the result that was obtained in Einstein gravity when $\Lambda$ was very small($\Lambda<10^{-16}$). On the other hand, our results show that, by increasing $\Lambda$, the maximum mass of white dwarf decreases. The maximum mass of white dwarfs with the cosmological constant are in the range $M_{max}\leq 1.41M_{\odot}$. According to our results in table~\ref{lstructure}, the cosmological constant has no significant effect on the properties of white dwarfs when it's value is less than $10^{-16}~\rm m^{-2}$.
 In order to conduct further investigations, in figures~\ref{Mrho} and~\ref{lmr}, we show the gravitational mass versus central mass density and radius, different lines represent different values of the cosmological constant. It's evident from figures~\ref{Mrho} and \ref{lmr} that the behavior of mass as a function of the central mass density (or radius) is highly sensitive to the variation of cosmological constant.
  Meanwhile, figure~\ref{lmr} reveals that larger the $\Lambda$ is, smaller the radius $R$ is, which means that we can regard $\Lambda$ as an external pressure which prevents the increasing mass of white dwarf. Interestingly, one can also see that initially R increases with $M$ and then decreasing with $M$ for $\Lambda>1\times10^{-17}$ in figure~\ref{lmr}, which indicates that the effect of Einstein-$\Lambda$ gravity is more sensitive to the low mass white dwarfs, their radius become smaller in Einstein-$\Lambda$ gravity.
 We also investigate the properties such as Schwarzschild radius, average density, compactness, gravitational redshift and dynamical stability of white dwarf in Einstein-$\Lambda$ gravity.

\begin{table}[tbp]
\centering
\begin{tabular}{|c c c c c c c|}
\hline
$\Lambda (\rm m^{-2})$&$M_{\rm max}(M_\odot)$&$R(\rm km)$&$R_{Sch}(\rm km)$& $\bar{\rho}(10^{11}\rm kg~m^{-3})$ & $\sigma(10^{-2})$&$z(10^{-2})$\\
\hline
$1.00\times10^{-18}$ & 1.41 & 1070 & 4.16 & 5.47 & 0.39 & 0.19\\
\hline
$1.00\times10^{-17}$ & 1.41 & 1070 & 4.16 & 5.47 & 0.39 & 0.19\\
\hline
$1.00\times10^{-16}$ & 1.41 & 1063 & 4.15 & 5.56 & 0.39 & 0.19\\
\hline
$1.00\times10^{-15}$ & 1.37 & 1009 & 4.04 & 6.31 & 0.40 & 0.18\\
\hline
$5.00\times10^{-15}$ & 1.24 & 877 & 3.65 & 8.70 & 0.42 & 0.14\\
\hline
$1.00\times10^{-14}$ & 1.12 & 790 & 3.30 & 10.74 & 0.42 & 0.10\\
\hline
$2.00\times10^{-14}$ & 0.94 & 692 & 2.79 & 13.54 & 0.40 & 0.04\\
\hline
$3.00\times10^{-14}$ & 0.82 & 631 & 2.42 & 15.52 & 0.38 & -0.007\\
\hline

\end{tabular}

\caption{\label{lstructure} Structure properties of white dwarf with different values of $\Lambda$, where $M_{\rm max}$, $R$, $R_{\rm Rsch}$, $\bar{\rho}$, $\sigma$,$z$ are Maximum mass, radius, Schwarzschild radius, average density and compactness of white dwarfs, respectively. The central density is $2\times 10^{13}~\rm kg/m^3$.}
\end{table}

\subsection{Schwarzschild radius}
The Schwarzschild radius of Einstein-$\Lambda$ gravity in the presence of the cosmological constant is~\cite{a,Bordbar2016}
\begin{equation}
R_{Sch}=\frac{\left[\left(\frac{3GM}{c^2}+\sqrt{\frac{9G^2M^2}{c^4}+\frac{1}{\Lambda}}\right)\right]^{\frac{1}{3}}}{\Lambda^{\frac{1}{3}}}-\frac{1}{\left[\Lambda^2\left(\frac{3GM}{c^2}+\sqrt{\frac{9G^2M^2}{c^4}+\frac{1}{\Lambda}}\right)\right]^{\frac{1}{3}}}.
\end{equation}
The results show that the Schwarzschild radius is a decreasing function of the cosmological constant when $\Lambda>1\times10^{-16}~\rm m^{-2}$, for $\Lambda<1\times10^{-16}~\rm m^{-2}$, the cosmological constant doesn't affect the Schwarzschild radius.

\subsection{Average density}
Now, using the effective mass and radius obtained in the Einstein-$\Lambda$ gravity, the average density of a white dwarf has the following form:
\begin{equation}
\bar{\rho}=\frac{3M}{4\pi R^3},
\end{equation}
The central density of the white dwarf in table~\ref{lstructure} is $2.0\times10^{13}~\rm kg/m^{3}$, the average density for white dwarfs in table~\ref{lstructure} is less than the central density.

\subsection{Compactness}
The compactness of a spherical object indicates the strength of gravity. It was defined as the ratio of Schwarzschild radius to the radius of object:
\begin{equation}
\sigma=R_{Sch}/R,
\end{equation}
We can see from table~\ref{lstructure}, $\sigma$ is almost same for these white dwarfs, which indicates that the strength of gravity is almost same for these white dwarfs.

\subsection{Gravitational redshift}
The gravitational redshift in Einstein-$\Lambda$ gravity is~\cite{a,Bordbar2016}:
\begin{equation}
z=\frac{1}{\sqrt{1+\frac{\Lambda R^2}{3}-\frac{2GM}{c^2R}}}-1,
\end{equation}
Because the radius of white dwarf is greater than the Schwarzschild radius, the gravitational redshift is very small (two orders of magnitude less than a neutron star, see Ref. ~\cite{a,Bordbar2016,panah} for more details ), the results can be found in the last column of table~\ref{lstructure}.
The results show that, the gravitational redshift is a decreasing function of the cosmological constant, it's notable that, when $\Lambda=3\times 10^{-14}~\rm m^{-2}$, the redshift becomes negative. As white dwarfs are compact objects, the gravitational redshift should be positive, therefore, we can give the cosmological constant an upper limit as $\Lambda<3\times10^{-14}~\rm m^{-2}$.

\subsection{Dynamical stability}
Chandrasekhar introduced the dynamical stability of the stellar model against the infinitesimal radial adiabatic perturbation~\cite{chandra1964}. The dynamical stability condition is satisfied when the adiabatic index is over than $\frac{4}{3}$ (i.e. $\gamma>\frac{4}{3}$) everywhere within the isotropic star, where $\gamma$ is defined in the following form:
\begin{equation}
\gamma=\frac{\rho c^2+P}{c^2P}\frac{dP}{d\rho}.
\end{equation}

\begin{figure}[tbp]
\centering 
\includegraphics[width=.65\textwidth]{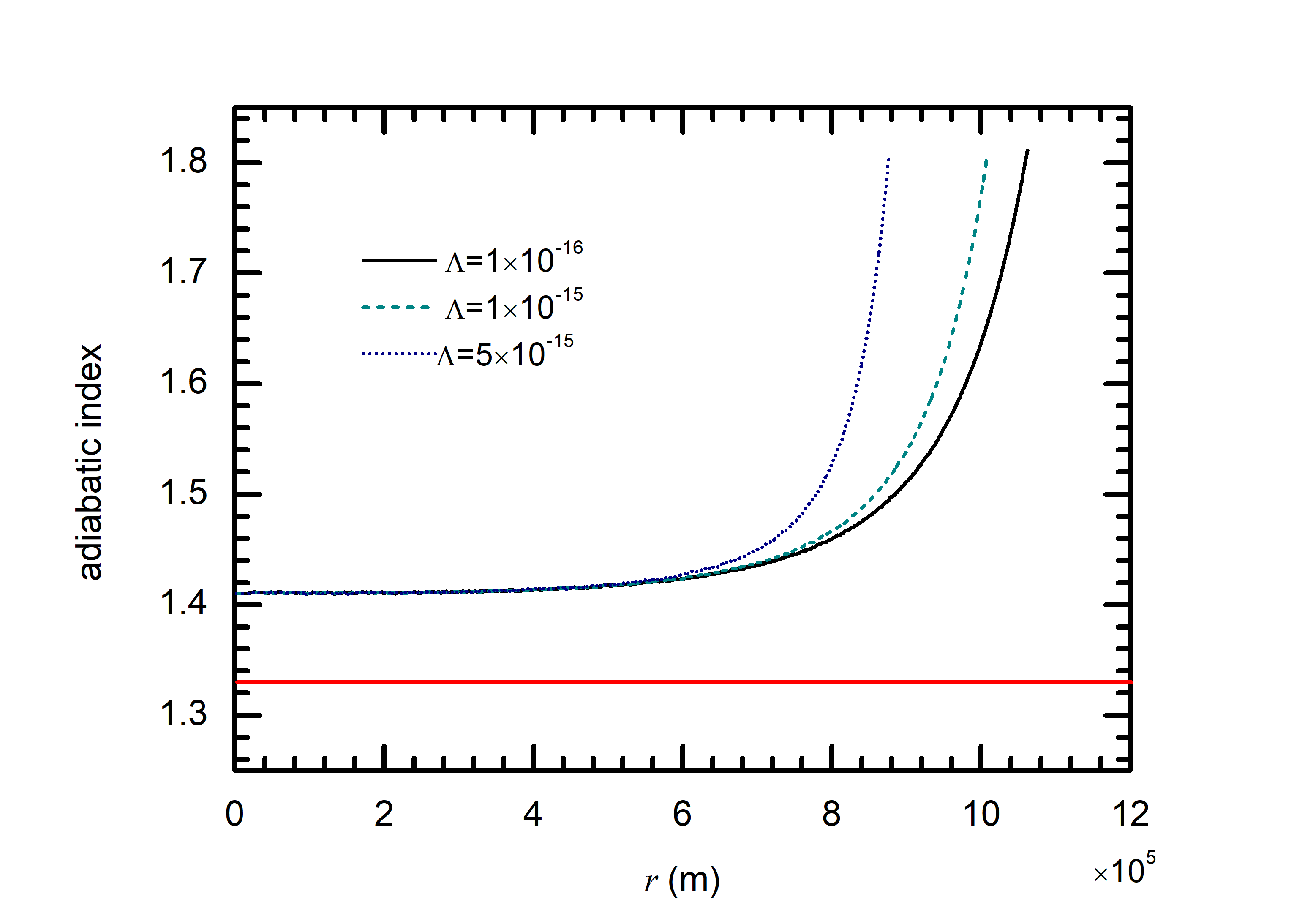}
\caption{\label{adiabatic index} Adiabatic index versus radius for $\Lambda=1.0\times10^{-16}$(solid), $\Lambda=1.0\times10^{-15}$(dash), $\Lambda=5.0\times10^{-15}$, where the central density is $2\times10^{13}~\rm kg/m^3$.}
\end{figure}

The stability condition had been developed and applied to astrophysical cases by many authors~\cite{bardeen,knutsen,mak,kalam}. Based on this stability condition, we investigate the adiabatic index versus the radius of white dwarfs in Einstein-$\Lambda$ gravity in figure~\ref{adiabatic index}. As one can see, these stars are dynamically stable.

\begin{figure}[tbp]
\centering 
\includegraphics[width=.3\textwidth]{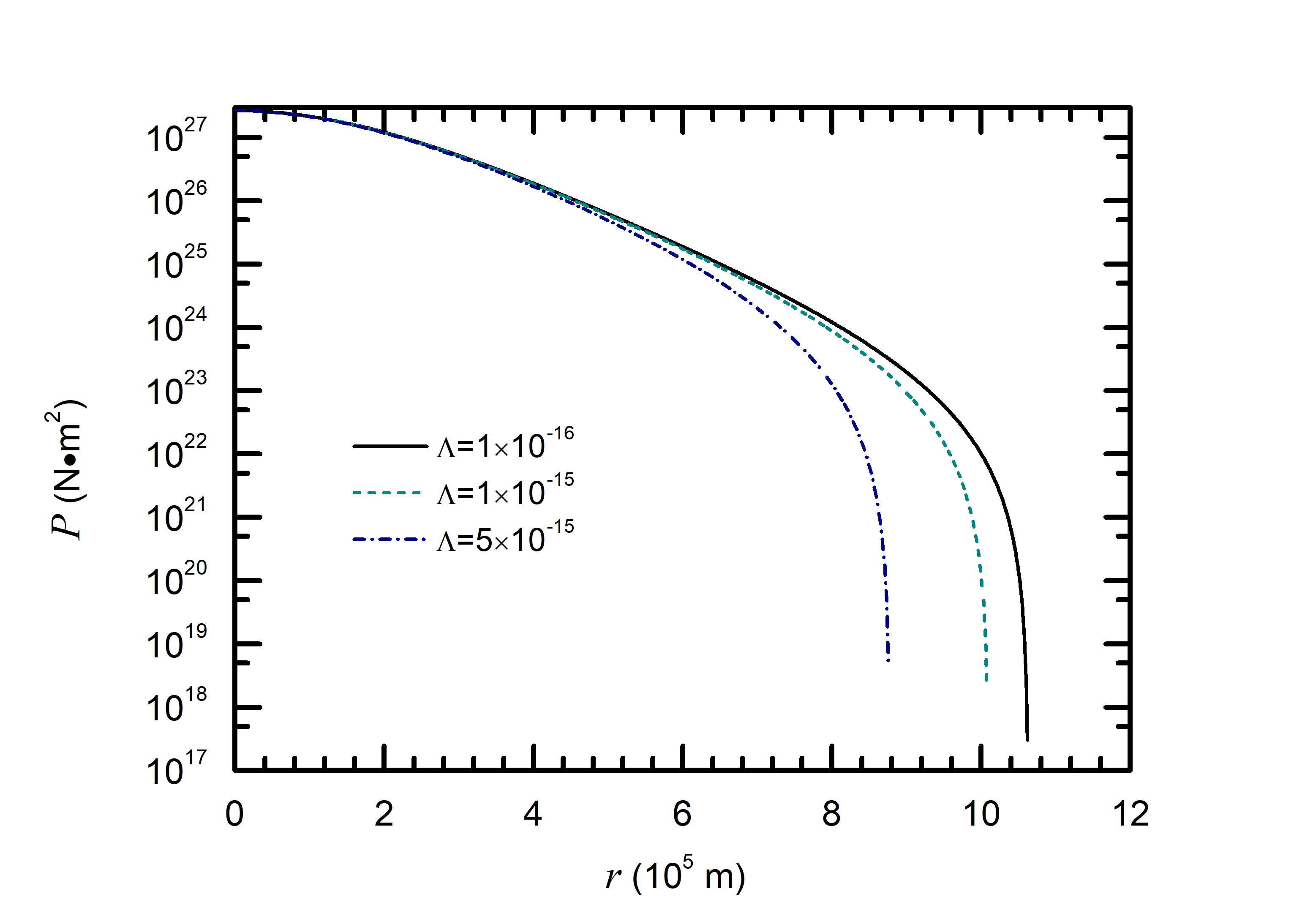}
\hfill
\includegraphics[width=.3\textwidth]{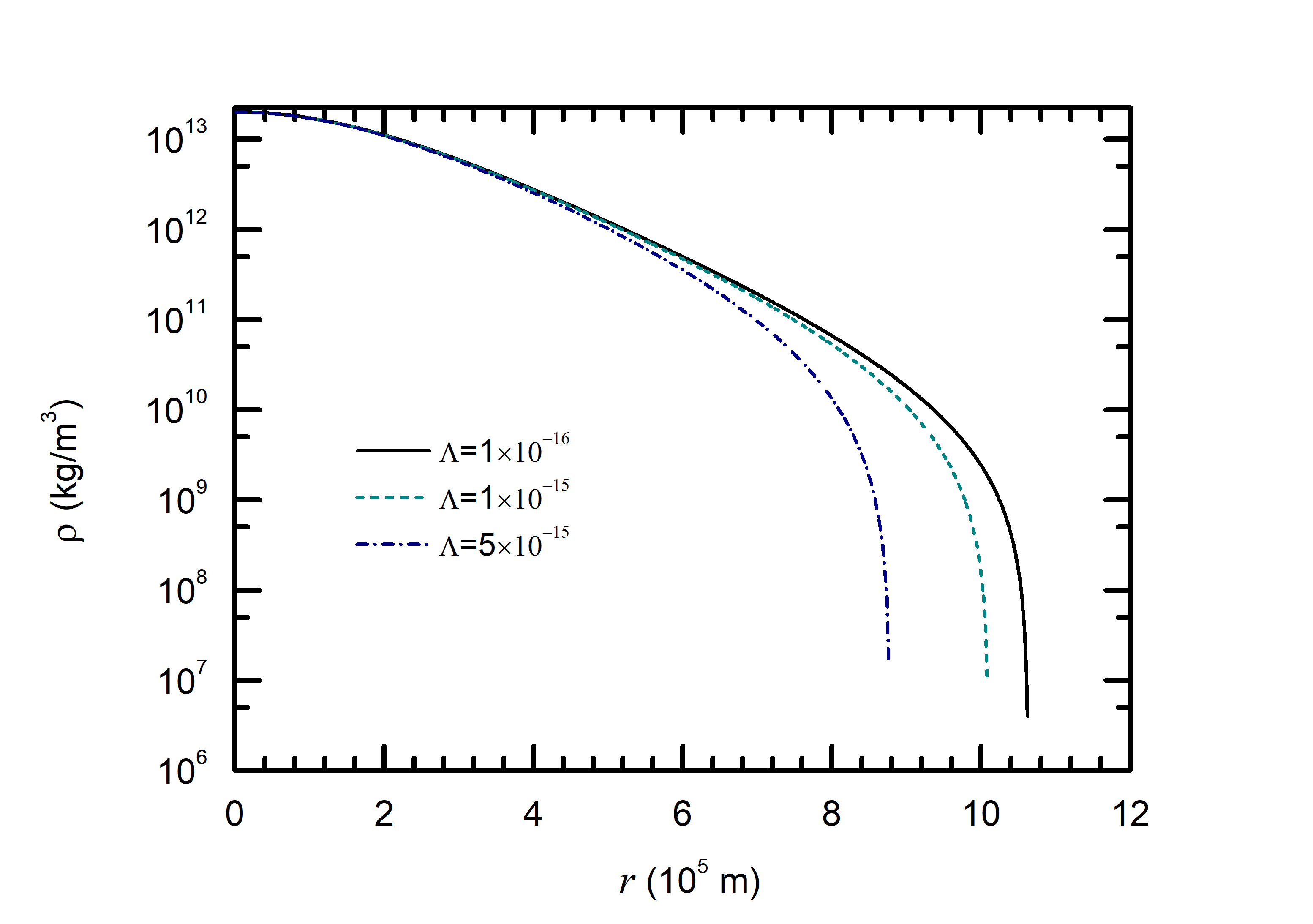}
\hfill
\includegraphics[width=.3\textwidth]{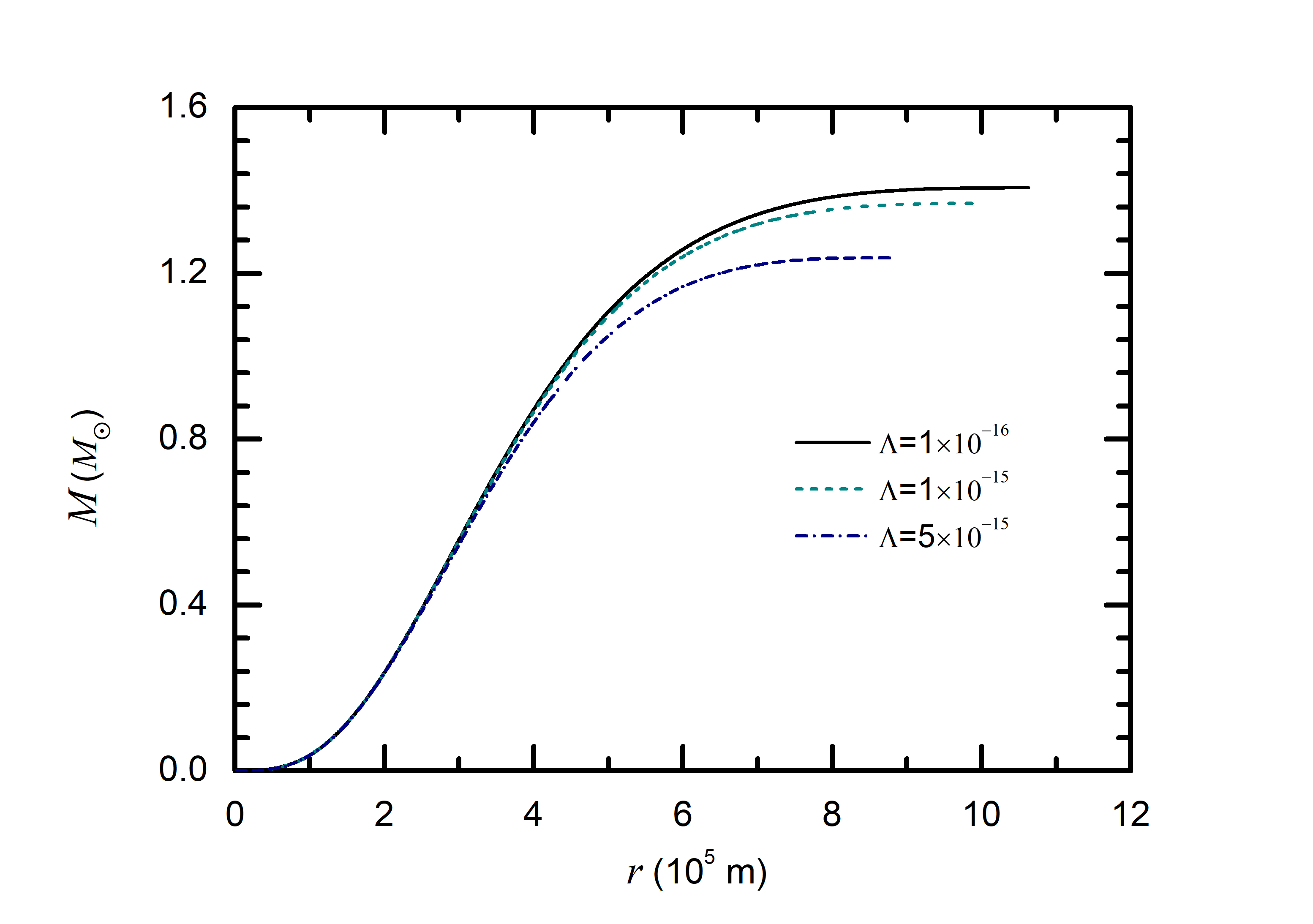}
\caption{\label{lprdr} Pressure (left), density (middle) and mass (right) versus radius for $\Lambda=1\times10^{-16}\rm $(solid), $\Lambda=1\times10^{-15}$(dash), $\Lambda=5\times10^{-15}$(dash dot), where the central density is set as $2\times10^{13}~\rm kg/m^3$.}
\end{figure}

In order to investigate internal structure of the white dwarf in the presence of cosmological constant in Einstein-$\Lambda$ gravity, we plot the pressure, density and mass versus distance from the center of a white dwarf in figure~\ref{lprdr}, in which the pressure and density are maximum at the center and decrease monotonically towards the boundary. One can also see clearly that the larger cosmological constant will make pressure and density decrease faster from the center to the surface. In the right panel, the mass increases monotonically towards the boundary, it is always positive within the star with different cosmological constants, for a given radius, the mass is decreasing as $\Lambda$ increase, which indicates that $\Lambda$  prevents the increasing mass of white dwarf.

\section{Conclusions}
In this paper, we discussed the properties of white dwarf with modified TOV equation in Einstein-$\Lambda$ gravity. We investigated the maximum mass of white dwarfs by using the Chandrasekhar equation of state of electron degenerate matter, our results showed that for $\Lambda<10^{-16}~\rm m^{-2}$, the cosmological constant did not affect the maximum mass and radius of white dwarfs. By increasing $\Lambda$, the maximum mass and radius of white dwarf are reduced.
Besides, we also investigated the effects of cosmological constant on the other properties of white dwarf such as: the Schwarzschild radius, average density, compactness, gravitational redshift and dynamical stability. The results indicated that the cosmological constant was a decreasing function of the maximum mass of white dwarfs when $\Lambda>10^{-16}\rm~m^{-2}$, for $\Lambda<10^{-16}\rm ~m^{-2}$, the cosmological constant didn't affect the properties of white dwarf (see table~\ref{lstructure} for more details). In addition, the gravitational redshift gave an upper limit of cosmological constant as $\Lambda<3\times10^{-14}\rm~m^{-2}$.  Also, our results showed that white dwarf in Einstein-$\Lambda$ gravity was dynamically stable.


\acknowledgments
We would like to thank anonymous referee for helpful and insightful comments. We wish to thank B. Eslam Panah for useful discussions.
This work has been supported by the National Natural Science Foundation of China under Nos. 11803026, 11473024, XinJiang University Science Fund, and the XinJiang Science Fund for Distinguished Young Scholars under No. QN2016YX0049.



\end{document}